\documentclass[12pt,a4paper]{article}

\usepackage[fleqn]{amsmath}
\usepackage{alltt}
\usepackage{amssymb}

\newtheorem{thm}{Theorem}
\newtheorem{lemma}[thm]{Lemma}

\author{Carl Bracken \\
School of Mathematical Sciences\\
University College Dublin\\
Ireland}

\def\Tr{{\rm Tr}}

\begin{document}

\title{New Families of Triple Error Correcting Codes with BCH Parameters}

\maketitle

\begin{abstract}
Discovered by  Bose, Chaudhuri and Hocquenghem \cite{BC}, \cite{H}, the BCH family of error correcting codes are one of the most studied families in coding theory. They are also among the best performing codes, particularly when the number of errors being corrected is small relative to the code length. In this article we consider binary codes with minimum distance of 7. We construct new families of codes with these BCH parameters via a generalisation of the Kasami-Welch Theorem.

\end{abstract}

\section{Introduction}Let $L=GF(2^n)$ for some odd $n$ and let $L^*$ denote  the non zero elements of $L$. Also, let $f(x)$ and $g(x)$ be two mappings from $L$ to itself. We assume the functions are chosen such that $f(0)=g(0)=0$. We can construct the parity check matrix $H$ of an error correcting code $C$ as follows.
 $$H =\left[\begin{array}{ccc} .... &   x   & ....\\
.... &   f(x)   & ....\\
.... &   g(x)   & ....\\
 \end{array} \right]$$
The code $C$ is defined as the nullspace of $H$. Each column is a binary vector of length $3n$ composed of the binary representations of three elements of $L^*$ with respect to some chosen basis. This matrix is a  $3n$ by $2^n-1$ array and hence $C$ has parameters $[2^n-1, 2^n-3n-1, d]$. This means that $C$ is a code of dimension $2^n-3n-1$ and minimum Hamming distance $d$ between any pair of vectors (usually called words). The number of errors a code can correct is given by $e=\frac{d-1}{2}$.  The code generated by the rows of $H$ is called the dual of $C$ and is denoted by $C^{\perp}$. We can determine the weights of the words in $C^{\perp}$ by computing the generalised Fourier transform of the pair of functions $\{f(x), g(x) \}.$ We define this transform as 
$$F^w(a,b,c)=\sum_{x \in GF(2^n)} (-1)^{Tr(ax+bf(x)+cg(x))}$$ 
with $b,c \in L^*$ and $a \in L$. In order to demonstrate that the minimum distance of $C$ is 7 we will require this transform to be limited to five specified values. We will also require at least one of our functions to be Almost Perfect Nonlinear (APN). The map $h$ is said to be APN on $L$ if the
number of solutions in $L$ of the equation
$$ h(x+q)-h(x)=p $$
is at most 2, for all $p\in L$ and $q \in L^*$.
An APN function can be used to construct a code with double error correcting BCH parameters.
For an introduction to the connections between Fourier transforms, APN functions and BCH codes, we recommend pages 1037-1039 of \cite{HB}.
The weights in $C^{\perp}$ are given by $w=\frac{1}{2}(2^n-V)$, where $V$ ranges over the values of $F^w(a,b,c)$.
If at least one of the pair $\{f(x), g(x) \}$ is APN then the code constructed with this pair will have a minimum distance of at least 5 as it will be the subcode of a code with double error correcting BCH parameters (\cite{HB} page 1037). The MacWilliams identities (\cite{HB} page 88) are a set of equations that allow us to compute the weight distribution of a linear code from the weight distribution of its dual. Let $A_w$ be the number of codewords in $C$ with weight $w$. As the distance in $C$ is at least 5 we have, $A_4 = A_3 = A_2 = A_1 = 0$ and $A_0 = 1$. If we demonstrate that the weights in $C^{\perp}$ take the same five values as those occuring in the dual of the triple error correcting BCH code, then we can use the MacWilliams identities to compute the multiplicity of the weights occuring in $C^{\perp}$. As we have five equations in five unknowns the multiplicity is exactly determined. Therefore $C^{\perp}$ and hence $C$ must have the same weight distribution as the triple error correcting BCH code. The words in the dual of the triple error correcting BCH take the five weights $2^{n-1}-2^{\frac{n-1}{2}}$, $2^{n-1}+2^{\frac{n-1}{2}}$, $2^{n-1}-2^{\frac{n+1}{2}}$, $2^{n-1}+2^{\frac{n+1}{2}}$ and $2^{n-1}$.  It follows that if we show that $F^w(a,b,c)$ is limited to the five values $0, \ \pm2^{\frac{n+1}{2}},\ \pm2^{\frac{n+3}{2}}$ for some pair of functions with at least one being APN, then the code constructed from this pair will have a minimum distance of 7.

\newpage

\section{Known Families of Codes}
The following table lists the known triple error correcting codes that can be constructed with pairs of functions over $GF(2^n)$ via the method outlined above.
\bigskip

\begin{center}

 \begin{tabular}{|c|c|c|}
 \hline
  &  &  \\
 ${\bf f(x), g(x) }$ & {\bf Conditions} & {\bf Ref. }\\
 \hline
  &  &  \\

 \small{$x^{2^k+1}$ , $x^{2^{2k}+1}$} & $gcd(n,k)=1$  & \small{\cite{BC}, \cite{H}} \\
& & Theorem 1 \\
 \hline
   &  & \\
 \small{$x^{2^k+1}$ , $x^{2^{3k}+1}$} &  $gcd(n,k)=1$  & \small{\cite{MS}}  \\ 
& & Theorem 1 \\
 \hline
  &  &  \\
 \small{$x^{2^{t}+1}$ , $x^{2^{t+2}+3}$} &  $n=2t+1$  & \small{\cite{TH}}\\
  &  &  \\
 \hline
  &  &  \\
 \small{$x^{2^{2k}-2^k+1}$ , $x^{2^{4k}-2^{3k}+2^{2k}-2^{k}+1}$} &   $gcd(n,k)=1$ & \small{ Theorem 2}  \\
  &  &  \\

 \hline
 \end{tabular}\ \ .
\end{center}

\bigskip

Next we will demonstrate that the first two pairs of functions from the above table indeed allow us to construct codes with a minimum distance of 7. We do this by computing the generalised Fourier transforms of each pair.
We believe that the results in this section are well known to those working in the area but we have not seen a proof in the litreture. The codes given by the first family are a generalisation of the classic, also called primitive, BCH codes defined by the pair of functions  $\{ x^3, x^5 \}$.
The second family generalises the non-BCH triple error correcting codes from \cite{MS} defined by the pair of functions $\{ x^{2^{t}+1}, x^{2^{t-1}+1} \}$ when $n=2t+1$. Before we proceed, we state the following often used result, a proof of which can be found in \cite{BBMM}.

\bigskip

\begin{lemma}
Let $L = GF(2^n)$ and let $p(x) = \sum_{i=0}^d r_i x^{2^{ki}}$ be a polynomial in $L[x]$ with $r_i \in L$ and $\gcd(k,n)=1$. Then  $p(x)$ has at most $2^d$ solutions in $L$. 
\end{lemma}

\bigskip

\begin{thm}
Let $n$ be odd and $k$ relatively prime to $n$. The pair of functions  
$\{x^{2^k+1}, g(x) \}$  construct a triple error correcting code with BCH parameters whenever
 $g(x)=x^{2^{2k}+1}$ or $x^{2^{3k}+1}$.
\end{thm}
Proof:\\
\noindent
As $x^{2^k+1}$ is APN \cite{G}, we can obtain the result by showing that the generalised Fourier transform always takes one of the five values $0, \ \pm2^{\frac{n+1}{2}}$ or $\pm2^{\frac{n+3}{2}}$.
Let $g(x)=x^{2^{tk}+1}$ where $t=2$ or $3$.\\
By definition, we have
$$  F^W(a,b,c) = \sum_{x \in L}
(-1)^{\Tr(ax + bx^{2^k+1} + cx^{2^{tk}+1})}. $$ 
We let $Q(x)=ax + bx^{2^k+1} + cx^{2^{tk}+1}$ and take the square of the transform to get,
$$  (F^W(a,b,c))^2   =  \sum_{x \in L}\sum_{y \in L}(-1)^{\Tr(Q(x))}(-1)^{\Tr(Q(y))}. $$
Replace $y$ with $x+u$ to obtain 
$$(F^w(a,b,c))^2=\sum_{u} (-1)^{Tr(Q(u))} \sum_{x} (-1)^{Tr(x(L(u)))},$$
where 
$$L(u)=bu^{2^{k}}+b^{2^{-k}}u^{2^{-k}}+cu^{2^{tk}}+c^{2^{-tk}}u^{2^{-tk}}.$$
Next we use the the fact that $ \sum_{x} (-1)^{Tr(\gamma x)} $ is $ 2^n $ when $ \gamma=0 $, and is $0$ otherwise to obtain
$$(F^w(a,b,c))^2=2^n \sum_{u \in K} (-1)^{Tr(Q(u))}$$
where $K$ is the kernel of $L(u).$
Consider $ \chi(u)=(-1)^{Tr(Q(u))}$. It can be easily demonstrated that $ \chi_a $ is a character of $K$ as $ \chi(u+v)=\chi(u)\chi(v).$
Now using the fact that, for a character $\chi $ of a group $H$,
$\sum_{h \in H} \chi(h)=  |H| $ if $\chi$ is the identity character and $0$ otherwise (see \cite{S}, 62-63) we see that $(F^w(a,b,c))^2=0$ or $2^{n+s}$, where $2^s$ is the number of solutions to $L(u)=0$.
Raising $L(u)$ by $2^{tk}$ we obtain the following equation
$$b^{2^{tk}}u^{2^{(t+1)k}}+b^{2^{(t-1)k}}u^{2^{(t-1)k}}+cu^{2^{2tk}}+cu=0.$$
Next we apply Lemma 1 and hence bound $s$ by $2^s \leq 2^4$ when $t=2$ and $2^s \leq 2^3$ when $t=3$. As 
$F^w(a,b,c)$ must be an integer, in both cases we have $F^w(a,b,c)=0$ , $\pm2^{\frac{n+1}{2}}$ or $\ \pm2^{\frac{n+3}{2}}$ and we are done.

\section{New Families of Codes}
In this section we prove that the fourth pair of functions from the list yield a code that is triple error correcting. Again, we do this by computing the generalised Fourier transform. The proof uses some of the techniques from Hans Dobbertin's proof of the Kasami-Welch theorem \cite{D}.
\begin{thm}
For $n$ odd and $k$ relatively prime to $n$, the pair of functions  
$\{x^{2^{2k}-2^k+1}, x^{2^{4k}-2^{3k}+2^{2k}-2^{k}+1} \}$ on $GF(2^n)$ construct a triple error correcting code with BCH parameters.
\end{thm}
Proof:\\
\noindent
Again, as $x^{2^{2k}-2^k+1}$ is APN \cite{D}, we achieve the result by computing the generalised Fourier transform.
By definition, \\

$F^w(a,b,c)=$ 
$$\sum_{x \in GF(2^n)} (-1)^{Tr(ax+bx^{2^{2k}-2^k+1}+cx^{2^{4k}-2^{3k}+2^{2k}-2^{k}+1})}.$$
Replacing $x$ with $x^{2^k+1}$ (a permutation) we obtain
$$F^w(a,b,c)=\sum_{x \in GF(2^n)} (-1)^{Tr(ax^{2^k+1}+bx^{2^{3k}+1}+cx^{2^{5k}+1})}.$$
Letting $Q(x)=ax^{2^k+1}+bx^{2^{3k}+1}+cx^{2^{5k}+1}$ and squaring gives,
$$(F^w(a,b,c))^2=\sum_{x}\sum_{y}(-1)^{Tr(Q(x)+Q(y))}.$$
Let $y=x+u$ to obtain
$$(F^w(a,b,c))^2=\sum_{u} (-1)^{Tr(Q(u))} \sum_{x} (-1)^{Tr(x(L(u)))}$$
where 
$$L(u)=au^{2^k}+a^{2^{-k}}u^{2^{-k}}+bu^{2^{3k}}+b^{2^{-3k}}u^{2^{-3k}}+cu^{2^{5k}}+c^{2^{-5k}}u^{2^{-5k}}.$$ 
As $ \sum_{x} (-1)^{Tr(\gamma x)} $ is $ 2^n $ when $ \gamma=0 $, and is $0$ otherwise, we have
$$(F^w(a,b,c))^2=2^n\sum_{u \in K} (-1)^{Tr(Q(u))}$$
where $K$ is the kernel of $L(u)$. Let $K=S_0 \cup S_1 $, where
$$S_0=\{u \in K :Tr(Q(u))=0\}$$
and
$$S_1=\{u \in K:Tr(Q(u))=1\}.$$
We now see that
$$ F^w(a,b,c)=\pm \sqrt{2^n( |S_0| - |S_1|}.$$
Therefore, we need only show that $ |S_0| - |S_1|  < 32$ for the result to follow. To this end we
let

\bigskip

$$G(u)=au^{2^k+1}+bu^{2^{3k}+1}+b^{2^{-k}}u^{2^{2k}+2^{-k}}+b^{2^{-2k}}u^{2^{k}+2^{-2k}}+cu^{2^{5k}+1}+$$
$$c^{2^{-k}}u^{2^{4k}+2^{-k}}+c^{2^{-2k}}u^{2^{3k}+2^{-2k}}+c^{2^{-3k}}u^{2^{2k}+2^{-3k}}+c^{2^{-4k}}u^{2^{k}+2^{-4k}}.$$
Note that $$G(u)+G(u)^{2^{-k}}=uL(u).$$
Therefore $u \in K $ if and only if $ G(u)=0 $ or $ 1 $ (as $gcd(k,n)=1$). Furthermore,
$u \in S_0 \ $if and only if $G(u)=0.$\\
Next let $ \chi(u)=(-1)^{Tr(au^{2^k+1}+bu^{2^{3k}+1}+cu^{2^{5k}+1})}$. As in Lemma 2, it can be shown that $ \chi_a $ is a character of $K$ and hence  $|S_0|-|S_1|=0 $ or $|K|$.
As $$|S_0| - |S_1|=\sum_{u\in K}\chi(u)$$
it follows that if $F^w(a) \neq 0 $ then $K=S_0$. That is we can assume $ L(u)=0 $ has the same
solution set as $ G(u)=0 $ and that  $|S_1|$ has no solutions when  $F^w(a,b,c) \neq 0$. Furthermore, as
$$ F^w(a,b,c) \in \{0, \pm \sqrt{2^n(|S_0|)} \}$$
we can say that $|S_0|$ must be an odd power of two.
It remains to show that $ G(u)=0 $ has at most 16 solutions when $K=S_0$.
For the sake of contradiction we assume  $ G(u)=0 $ has 32 solutions and that they form an additive group.
Now apply the identity
$$(u+v)(vG(u)+uG(v))+uv(G(u+v))=0  \hspace{1cm} (1) $$
for three solutions $u, v $ and $ u+v $ with $u \neq v$ and $v \neq 0$. We then rearrange to obtain the following expression
$$ b^{2^{-k}}(u^{2^{2k}}v+uv^{2^{2k}})(u^{2^{-k}}v+uv^{2^{-k}})+b^{2^{-2k}}(u^{2^k}v+uv^{2^k})(u^{2^{-2k}}v+uv^{2^{-2k}})+$$
$$c^{2^{-k}}(u^{2^{4k}}v+uv^{2^{4k}})(u^{2^{-k}}v+uv^{2^{-k}})+c^{2^{-2k}}(u^{2^{3k}}v+uv^{2^{3k}})(u^{2^{-2k}}v+uv^{2^{-2k}})+$$
$$c^{2^{-3k}}(u^{2^{2k}}v+uv^{2^{2k}})(u^{2^{-3k}}v+uv^{2^{-3k}})+c^{2^{-4k}}(u^{2^{k}}v+uv^{2^{k}})(u^{2^{-4k}}v+uv^{2^{-4k}}) =0.$$
Under our assumption the above equation has 32 solutions in $u$, for a fixed (non zero) solution $v$.
Next let $u=vw$ and divide by $v^2$ to obtain
$$ b^{2^{-k}}v^{2^{2k}+2^{-k}}(w+w^{2^{2k}})(w+w^{2^{-k}})+b^{2^{-2k}}v^{2^{k}+2^{-2k}}(w+w^{2^k})(w+w^{2^{-2k}})+$$
$$c^{2^{-k}}v^{2^{4k}+2^{-k}}(w+w^{2^{4k}})(w+w^{2^{-k}})+c^{2^{-2k}}v^{2^{3k}+2^{-2k}}(w+w^{2^{3k}})(w+w^{2^{-2k}})+$$ 
$$c^{2^{-3k}}v^{2^{2k}+2^{-3k}}(w+w^{2^{2k}})(w+w^{2^{-3k}})+c^{2^{-4k}}v^{2^{k}+2^{-4k}}(w+w^{2^{k}})(w+w^{2^{-4k}})=0.$$
Now letting $w+w^{2^{-k}}=r$, we obtain the following equation, with 16 solutions in $r$ which form an additive group as $w+w^{2^{-k}}$ is $2-$to$-1$ and linear.
$$b^{2^{-k}}v^{2^{2k}+2^{-k}}(r^{2^k}+r^{2^{2k}})(r)+b^{2^{-2k}}v^{2^{k}+2^{-2k}}(r^{2^{-k}})(r+r^{2^{-k}})+$$
$$c^{2^{-k}}v^{2^{4k}+2^{-k}}(r^{2^k}+r^{2^{2k}}+r^{2^{3k}}+r^{2^{4k}})(r)+c^{2^{-2k}}v^{2^{3k}+2^{-2k}}(r^{2^k}+r^{2^{2k}}+r^{2^{3k}})(r+r^{2^{-k}})+$$
$$c^{2^{-3k}}v^{2^{2k}+2^{-3k}}(r^{2^k}+r^{2^{2k}})(r+r^{2^{-k}}+r^{2^{-2k}})+$$
$$c^{2^{-4k}}v^{2^{k}+2^{-4k}}(r^{2^k})(r+r^{2^{-k}}+r^{2^{-2k}}+r^{2^{-3k}})=0.$$
Rearranging gives
$$Ar^{2^{k}+1}+Br^{2^{2k}+1}+B^{2^{-k}}r^{2^{k}+2^{-k}}+Cr^{2^{3k}+1}+C^{2^{-k}}r^{2^{2k}+2^{-k}}+$$
$$C^{2^{-2k}}r^{2^{k}+2^{-2k}}+Dr^{2^{4k}+1}+D^{2^{-k}}r^{2^{3k}+2^{-k}}+D^{2^{-2k}}r^{2^{2k}+2^{-2k}}+D^{2^{-3k}}r^{2^{k}+2^{-3k}}=0, $$
where $A, B, C $ and $D$ are functions of $b,c$ and $v$, that is, they are fixed elements of the field.
Now, re-apply the identity $(1)$ for the above equation with three solutions $r, s$ and $r+s$ (where $s$ will be a
fixed non zero solution) to obtain
$$B^{2^{-k}}(r^{2^k}s+rs^{2^k})(r^{2^{-k}}s+rs^{2^{-k}})+C^{2^{-k}}(r^{2^{2k}}s+rs^{2^{2k}})(r^{2^{-k}}s+rs^{2^{-k}})+$$ 
$$C^{2^{-2k}}(r^{2^{k}}s+rs^{2^{k}})(r^{2^{-2k}}s+rs^{2^{-2k}})+D^{2^{-k}}(r^{2^{3k}}s+rs^{2^{3k}})(r^{2^{-k}}s+rs^{2^{-k}})+$$
$$D^{2^{-2k}}(r^{2^{2k}}s+rs^{2^{2k}})(r^{2^{-2k}}s+rs^{2^{-2k}})+D^{2^{-3k}}(r^{2^{k}}s+rs^{2^{k}})(r^{2^{-3k}}s+rs^{2^{-3k}})=0.$$
Next let $r=st$, to give the following equation which will have 16 solutions in $t$:
$$B^{2^{-k}}s^{2^{k}+2^{-k}}(t+t^{2^{k}})(t+t^{2^{-k}})+C^{2^{-k}}s^{2^{2k}+2^{-k}}(t+t^{2^{2k}})(t+t^{2^{-k}})+$$
$$C^{2^{-2k}}s^{2^{k}+2^{-2k}}(t+t^{2^{k}})(t+t^{2^{-2k}})+D^{2^{-k}}s^{2^{3k}+2^{-k}}(t+t^{2^{3k}})(t+t^{2^{-k}})+$$ 
$$D^{2^{-2k}}s^{2^{2k}+2^{-2k}}(t+t^{2^{2k}})(t+t^{2^{-2k}})+D^{2^{-3k}}s^{2^{k}+2^{-3k}}(t+t^{2^{k}})(t+t^{2^{-3k}})=0.$$

\newpage

Now let $t+t^{2^{-k}}=z$ to obtain
$$B^{2^{-k}}s^{2^{k}+2^{-k}}(z^{2^{k}})(z)+C^{2^{-k}}s^{2^{2k}+2^{-k}}(z^{2^{k}}+z^{2^{2k}})(z)+$$
$$C^{2^{-2k}}s^{2^{k}+2^{-2k}}(z^{2^{k}})(z+z^{2^{-k}})+D^{2^{-k}}s^{2^{3k}+2^{-k}}(z^{2^{k}}+z^{2^{2k}}+z^{2^{3k}})(z)+$$
$$D^{2^{-2k}}s^{2^{2k}+2^{-2k}}(z^{2^{k}}+z^{2^{2k}})(z+z^{2^{-k}})+D^{2^{-3k}}s^{2^{k}+2^{-3k}}(z^{2^{k}})(z+z^{2^{-k}}+z^{2^{-2k}})=0.$$
The above equation will have eight solutions in $z$, which will again form an additive group. Now apply the identity $(1)$ for three solutions $z, p$ and $z+p$ to get
$$E(z^{2^k}p+zp^{2^k})(z^{2^{-k}}p+zp^{2^{-k}})+F(z^{2^{2k}}p+zp^{2^{2k}})(z^{2^{-k}}p+zp^{2^{-k}})+$$ $$F^{2^{-k}}(z^{2^k}p+zp^{2^k})(z^{2^{-2k}}z+zp^{2^{-2k}})=0,$$
where $E$ and $F$ are fixed elements of the field. Letting $q=zp$ and dividing by $p^2$ yields
$$Ep^{2^{k}+2^{-k}}(q+q^{2^{k}})(q+q^{2^{-k}})+Fp^{2^{2k}+2^{-k}}(q+q^{2^{2k}})(q+q^{2^{-k}})+$$
$$F^{2^{-k}}p^{2^{k}+2^{-2k}}(q+q^{2^{k}})(q+q^{2^{-2k}})=0.$$
Now let $q+q^{2^{-k}}=d$, to obtain the following equation with four solutions in $d$:
$$Ep^{2^{k}+2^{-k}}(d^{2^{k}})(d)+Fp^{2^{2k}+2^{-k}}(d^{2^{k}}+d^{2^{2k}})(d)+F^{2^{-k}}p^{2^{k}+2^{-2k}}(d^{2^{k}})(d+d^{2^{-k}})=0.$$
Apply the identity $(1)$ once more, for three solutions $d, e$ and $d+e$ to obtain
$$F^{2^{-k}}p^{2^{k}+2^{-2k}}(d^{2^k}e+de^{2^k})(d^{2^{-k}}e+de^{2^{-k}})=0.  $$
It needs to be verified at this point that the constant term $F^{2^{-k}}p^{2^{k}+2^{-2k}}$ is non-zero. This is true as the term is the product of a collection of non-zero terms. It follows that the above equation implies $(d^{2^{-k}}e+de^{2^{-k}})^{2^{k}+1}=0$, which gives $d^{2^{-k}}e+de^{2^{-k}}=0$ for which the only solution is $d=e$.
This contradiction completes the proof.

\end{document}